# Polariton Nanophotonics using Phase Change Materials

Kundan Chaudhary[1,‡], Michele Tamagnone[1,‡,*], Xinghui Yin[1,‡,*], Christina M. Spägele[1,‡], Stefano L. Oscurato[1,2], Jiahan Li[3], Christoph Persch[4], Ruoping Li[1], Noah A. Rubin[1], Luis A. Jauregui[5], Kenji Watanabe[6], Takashi Taniguchi[6], Philip Kim[5], Matthias Wuttig[4], James H. Edgar[3], Antonio Ambrosio[7], Federico Capasso[1,*]

[1] Harvard John A. Paulson School of Engineering and Applied Sciences, Harvard University, Cambridge, Massachusetts 02138, USA.

[2] Department of Physics "E. Pancini", University of Naples "Federico II", Complesso Universitario di Monte S. Angelo, Via Cinthia 21, 80126, Naples, Italy.

[3] Department of Chemical Engineering, Kansas State University, Manhattan, Kansas 66506, USA.

[4] 1. Physikalisches Institut IA, RWTH Aachen University, 52056 Aachen, Germany.

[5] Department of Physics, Harvard University, Cambridge, Massachusetts 02138, USA.

[6] National Institute for Materials Science, 1-1 Namiki, Tsukuba, 305-0044, Japan

[7] Center for Nanoscale Systems, Harvard University, Cambridge, Massachusetts 02138, USA.

*Correspondence to mtamagnone@seas.harvard.edu, xyin@seas.harvard.edu, capasso@seas.harvard.edu

‡These authors contributed equally to this work.

**Polaritons formed by the coupling of light and material excitations such as plasmons, phonons, or excitons enable light-matter interactions at the nanoscale beyond what is currently possible with conventional optics[1-18]. Recently, significant interest has been attracted by polaritons in van der Waals materials, which could lead to applications in sensing,[15] integrated photonic circuits[4] and detectors[3]. However, novel techniques are required to control the propagation of polaritons at the nanoscale and to implement the first practical devices. Here we report the experimental realization of polariton refractive and meta-optics in the mid-infrared by exploiting the properties of low-loss phonon polaritons in isotopically pure hexagonal boron nitride (hBN)[5], which allow it to interact with the surrounding dielectric environment comprising the low-loss phase change material, $Ge_3Sb_2Te_6$ (GST)[19-28]. We demonstrate waveguides which confine polaritons in a 1D geometry, and refractive optical elements such as lenses and prisms for phonon polaritons in hBN, which we characterize using scanning near field optical microscopy. Furthermore, we demonstrate metalenses, which allow for polariton wavefront engineering and sub-wavelength focusing. Our method, due to its sub-diffraction and planar nature, will enable the realization of programmable miniaturized integrated optoelectronic devices, and will lay the foundation for on-demand biosensors.**

Phonon polaritons (PhP) in thin films of hBN behave as confined guided optical modes, which extend as evanescent waves into the semi-spaces above and below the hBN. Therefore, the propagation of PhPs is affected by the refractive indices of the superstrate and substrate[3,4,6,8]. These are not interface modes but exist inside the volume of hBN. The permittivity values are of opposite signs along different principal axes and thus polaritons exhibit hyperbolic dispersion[3,8,12]. The degree to which the optical energy density of polaritons extends into the substrate and superstrate depends on the wavelength and thickness of hBN. Therefore, the excitation wavelength can be controlled to the point where the polariton is affected even by the very first few nanometres of the substrate and superstrate[17]. This suggests the feasibility of substrate-engineered polariton optics where, instead of nanopatterning the polaritonic material itself, optical functions such as waveguiding and focusing are conferred through engineering the refractive index of the substrate.

A heterostructure comprising the phase change material $Ge_3Sb_2Te_6$ (GST) and isotopically pure $h^{11}BN$ (referred to hBN hereinafter) is the ideal system for a proof-of-concept demonstration of substrate engineered polariton optics: hBN possess low-loss polaritons with large propagation lengths[5] and GST can support two very different refractive indices in its amorphous and crystalline phases ($n_a = 4.2$ and $n_c = 6.1$), which can co-exist at room temperature. Additionally, GST allows for reconfigurability in terms of writing, erasing, and re-writing optical components because it can be reversibly switched between its two phases through electrical or optical stimuli. This phenomenon is used in commercial rewritable optical discs[21].

While GST has been previously employed to demonstrate switchable phonon polariton resonators in quartz[26], more complex applications including metasurfaces remained elusive due to limited propagation lengths of phonon-polaritons in quartz. On the other hand, $VO_2$ has been used as a substrate with hBN to achieve temperature dependent polariton dispersion[27]. However, this approach suffers from limitations because $VO_2$'s different phases cannot co-exist at the same temperature, and thus in close spatial proximity as is desirable for the realisation of optical devices.

Here, we show an hBN-GST heterostructure in which arbitrary patterns can be written, erased, and re-written to control the PhP propagation. We achieve this by defining several structures, ranging from waveguides to diffraction-limited focusing metalenses. Specifically, we use low-loss PhPs in isotopically pure hBN ($^{11}B$ isotopes with >99% purity[5]) with longer propagation lengths, which we combine with $Ge_3Sb_2Te_6$, a GST stoichiometry with particularly low absorption in the mid-infrared (mid-IR)[28]. We work in the second Reststrahlen band (from 1361-1610 cm$^{-1}$), which is associated with in-plane optical phonons[5,6].

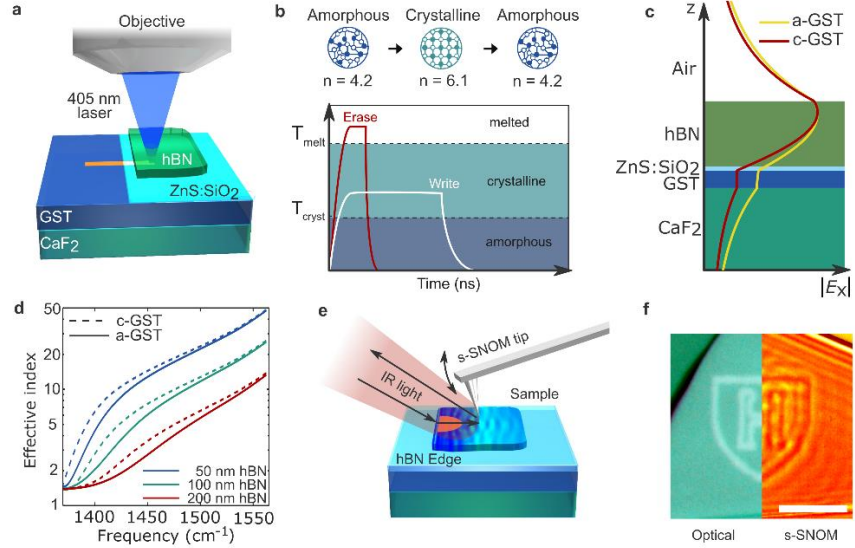

**Figure 1 | Reconfigurable polaritons in hBN-GST heterostructures**. **a,** Writing setup and device cross-section. A 405 nm focused laser beam is used to write and reconfigure devices on GST underneath hBN (transparent at 405 nm). **b,** Longer, low power laser pulses are used to crystallise GST and shorter high-power pulses are used to melt it to restore the amorphous phase. **c,** Electric field profile of polaritons for the a-GST and c-GST cases. The electric field confinement is larger in c-GST (due to its larger refractive index) than in a-GST. $E_x$ represents the electric field along the direction of polariton propagation. Thicknesses for each layer are 195 nm for hBN, 15 nm for ZnS:SiO$_2$, 55 nm for GST and 1 mm for CaF$_2$, which is then considered semi-infinite. Refractive indices are 1.7 for ZnS:SiO$_2$, 4.2 and 6.1 for GST in amorphous and crystalline phases, respectively, 1.37 for CaF$_2$ while hBN is modelled with the Lorentz model presented in the Supplementary Information. **d,** Calculated dispersion relation of the effective index $n_{\text{eff}}$ for different hBN thicknesses on a-GST and c-GST. **e,** Polaritons are launched by the hBN edge when light impinges on the sample. The launched polaritons interact with the written devices and their propagation is imaged using scattering-type scanning near field optical microscopy (s-SNOM). **f,** Example of optical and s-SNOM images. Scale bar is 5 µm.

As PhPs are strongly confined, a thin layer of 55 nm of GST below hBN (195 nm of thickness) is sufficient to significantly alter polariton propagation. To create the heterostructure, we sputter a thin film of GST on a CaF$_2$ substrate (in an amorphous phase as-deposited), protect it with a 15 nm layer of ZnS:SiO$_2$ against oxidation and then transfer exfoliated hBN to form the top layer. A pulsed laser diode is used to write and erase patterns by selectively crystallising or re-amorphising GST[21].

The PhP mode profile is affected by the state (either crystalline or amorphous) of the GST beneath it (Figure 1c), which can be quantified by using the effective index $n_{\text{eff}} = c/v_{\text{ph}}$, where $c$ is the speed of light and $v_\mathrm{ph}$ is the PhP's phase velocity. Figure 1d shows the dispersion of $n_\mathrm{eff}$ for different hBN thicknesses on both a-GST ($n_\mathrm{eff,a}$) and c-GST ($n_\mathrm{eff,c}$) respectively. We use scattering-type scanning near field optical microscopy (s-SNOM) to characterize the polaritons launched at hBN edges which propagate across the optical elements (Figure 1e,1f) [8,13,17].

Polariton propagation in heterostructures with a- or c-GST is analogous to light propagation in two different materials (such as air and glass). The continuity of the electric field at the boundary between two regions implies Snell's law[30]:

$$\frac{n_\mathrm{eff,c}}{n_\mathrm{eff,a}} = \frac{\sin(\theta_\mathrm{a})}{\sin(\theta_\mathrm{c})} \qquad (1)$$

where $\theta$ is the propagation angle in corresponding regions with respect to the interface normal.

Many conventional optical devices (such as lenses and prisms) are governed by Snell's law, suggesting that similar components can be implemented in our hBN-GST heterostructure. The first example to illustrate this principle is a refractive lens, specifically, a plano-convex semi-circular lens to focus PhPs (Figure 2a).

We write and erase a semi-circular plano-convex lens (radius ($R$) = 5 $\mu$m) twice and subsequently replace it with a plano-concave lens of the same radius (Figure 2b-f). The straight hBN edge launches linear waves (the planar equivalent of three-dimensional plane waves), which are either focused by the plano-convex lens or diverged by the plano-concave lens (Figure 1g,h). The lateral size of the focal spot is 2 $\mu$m (29 % of the free space wavelength), which is diffraction-limited according to the Abbe limit computed for the 2D waves (i.e., 2.08 $\mu$m). The numerical aperture (NA) with respect to polaritons in a-GST is 0.55, while the NA with respect to vacuum is 2.11, which is higher than unity due to the high-confinement of PhPs.

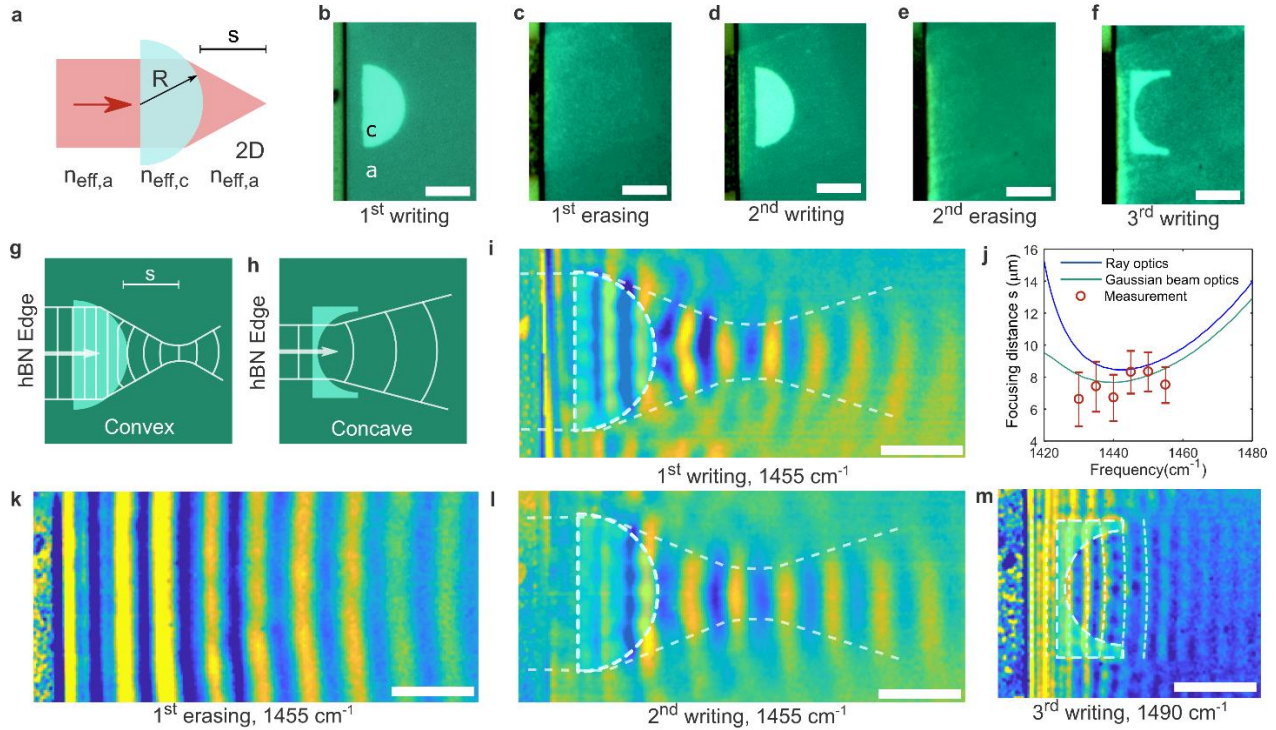

**Figure 2 | Rewritable flat polaritonic lenses**. **a,** Plano-convex lens schematics for 3D and 2D semi-spherical and semi-circular lenses. In the 2D case the material refractive index is replaced by the effective index of the polaritons on amorphous or crystalline GST. **b-f,** Optical images of the written lens. The written patterns are clearly visible in the pictures because the refractive index of a- and c-GST also differs at visible wavelengths. First a plano-convex semi-circular lens (radius R = 5 µm) is written and measured, then it is erased, rewritten (with same dimensions), erased again and finally the same area is reconfigured into a plano-concave lens (R = 5 µm). **g-h,** Diagram of wavefronts for 2D plano-convex and plano-concave lenses respectively. **i,** s-SNOM image of the plano-convex lens after the first writing. A focal spot is clearly visible. **j,** Dependence of the focal length on the wavenumber. **k,** s-SNOM scan after first erasing. **l,** s-SNOM scan of the re-written plano-convex lens. **m,** s-SNOM scan of the plano-concave lens (third writing). s-SNOM images in panels **i, l** and **m** have been processed to remove the fringes outside the main beam, see Supplementary Information for details. Scale bars are 5 µm.

We performed phase resolved s-SNOM measurements after each writing and erasing step. Using amplitude and phase information, the wavefronts of the polaritons can be clearly imaged (see Supplementary Information for more details on measurements and image processing). The resulting images confirm focusing, which is shown by the narrow waist in the transmitted beam (Figure 1i). Characteristic curved wavefronts can be seen before and after the focal spot. The position of the focal spot of the lens is measured from the images and compared with theoretical computation with two different methods (Figure 2j). The first method is based on computing the focal spot using ray optics, while the second, more accurate, method models the focused beam as a Gaussian beam, and identifies the focal spot as the beam-waist[29] (see more details in the Supplementary Information).

After erasing, only polaritons with linear wavefronts that are launched by the hBN edge are visible, whereas focusing is fully restored when the lens is rewritten (Figure 2k,l). Furthermore, reconfiguring the same area to a plano-concave lens results in curved wavefronts associated with diverging PhPs (Figure 2m).

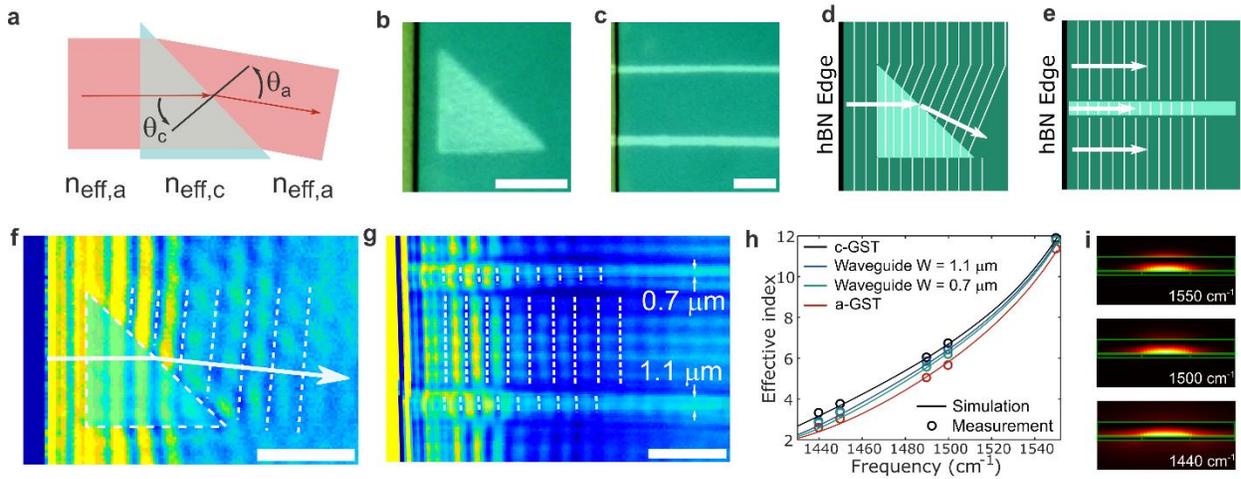

**Figure 3 | Prism and waveguides. a,** Snell's law for 2D prisms determines deflection of polaritons. **b,** Optical image of the written prism, an isosceles right-angled triangle with edges of 7.5 µm. The flake edge is also visible. **c,** Optical image of the written waveguides (top 0.7 µm wide, bottom 1.1 µm wide). The distance between the waveguides is 8.5 µm, which ensures no coupling between them. **d,** Diagram of wavefronts for the prism. **e,** Schematics of wavefronts for a waveguide. Polaritons propagating inside the waveguide have smaller fringe spacing due to the additional confinement of the waveguide mode. **f,** s-SNOM image of prism showing a clear deflection angle of the outgoing wavefronts. **g,** s-SNOM image of waveguides, showing the expected confinement of the modes inside of them. The fringe spacings are different for waveguides with different widths, confirming that the spacing is determined by the mode of the waveguide. **h,** Simulated and measured effective indices of the waveguides. The effective indices are between $n_{\text{eff,a}}$ and $n_{\text{eff,c}}$. **i,** Cross-section of the guided mode of the 0.7 µm waveguide at different frequencies (out-of-plane Poynting vector). Scale bars are 5 µm.

The successful implementation of lenses can be extended to other common devices such as prisms and waveguides (Figure 3). Planar prisms are simply triangles and follow Snell's law (Figure 3a). We wrote a prism and two waveguides (with different widths) close to the hBN edge so that edge-launched waves can interact with them (Figure 3b-3e). The prism is designed to be an isosceles right triangle with one side parallel to the hBN edge such that edge-launched waves enter orthogonal to it. When traversing the hypotenuse, the polariton propagation direction (*k*-vector) is bent downwards (as expected from Snell's law), as is clearly visible in the s-SNOM measurements in the form of bent fringes (Figure 3f).

The waveguides consist of c-GST lines with widths (0.7 and 1.1 µm) smaller or comparable to the guided polariton wavelength. They provide additional in-plane confinement such that the propagating mode is truly one-dimensional and is confined along the waveguide. Here, the c-GST line acts as the waveguide core, while a-GST serves as cladding. The s-SNOM measurement in Figure 3g shows that the wavefront spacing decreases inside the waveguides, as expected from confined modes. Furthermore, the compression is greater for the wider waveguide. This implies that the waveguide effective index $n_{\text{eff,wg}}$ is larger when the width of the waveguide increases, which agrees with the behaviour known from conventional dielectric waveguides where the core size affects the effective index of the mode. In both the conventional and the polariton cases, the value of the waveguide effective index lies between the indices of the core and cladding material, i.e. $n_{\text{eff,a}} \leq n_{\text{eff,wg}} \leq n_{\text{eff,c}}$. We verified this behaviour by numerically calculating the waveguide dispersion relation (see Methods) and comparing the results to s-SNOM measurements taken at different frequencies (Figure 3h). Figure 3i shows a cross-section of a guided mode obtained from numerical simulation, illustrating how polaritons are confined both vertically and laterally.

Metasurfaces have recently emerged as a novel and versatile method for engineering light propagation by using arrays of discrete elements, which locally alter the phase of transmitted light. By changing the size and shape of these elements, arbitrary predetermined phase profiles can be implemented[30]. Figure 4 shows the adaptation of this concept for polaritons and its implementation in an hBN-GST heterostructure. Our approach allows designing one-dimensional metalenses, which focus polaritons that propagate in two dimensions. The metalens elements are truncated c-GST waveguides defined in a-GST environment and create the hyperbolic phase profile[31]:

$$\phi(y) = -\frac{2\pi}{\lambda_{\text{eff,a}}}\left(\sqrt{y^2 + f^2} - f\right) \qquad (2)$$

where *y* is the element position in y-direction and f is the focal *length* (Figure 4a). We build a phase library for elements of varying lengths (Metalens 1, periodicity of unit cell: 1.8 µm) and widths (Metalens 2, periodicity of unit cell: 1.2 µm) (Figure 4b and c), and subsequently incorporate the required phase profile by choosing the corresponding elements (Figure 4d) [30]. First, we demonstrate a metalens based on length changes, then we erase and replace it with a metalens where only the width of the individual elements varies.

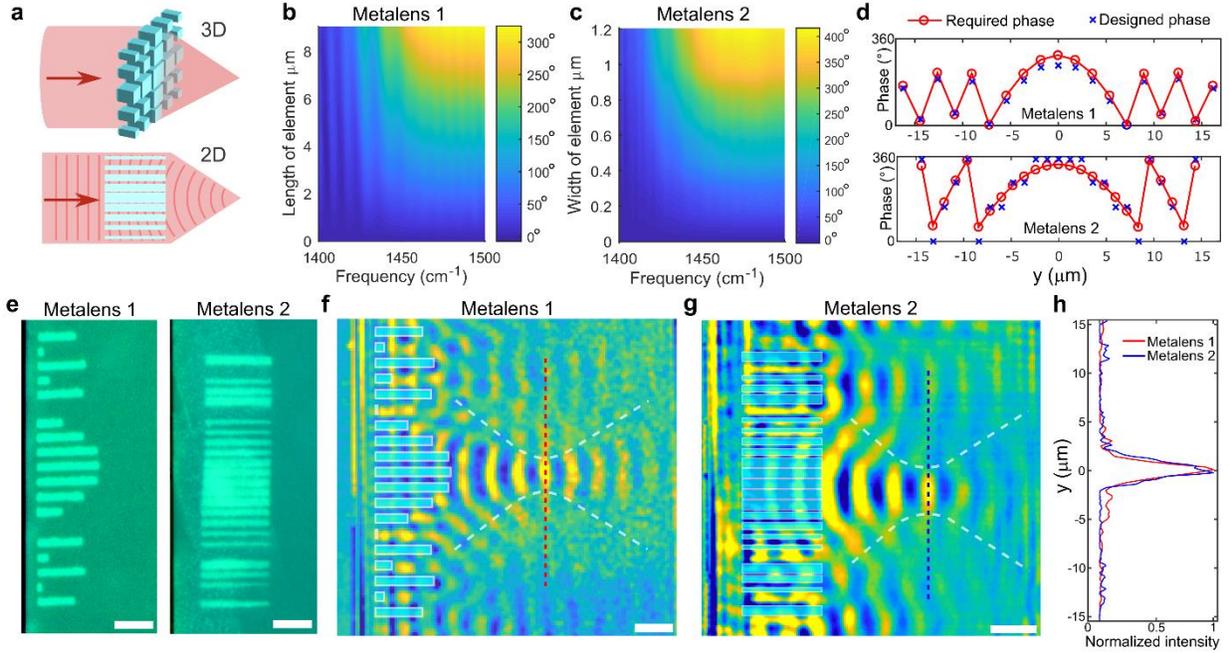

**Figure 4 | Reconfigurable Metalenses**. **a,** Adaptation of metalenses to the 2D case. Each discrete element changes the local phase of light, so the wavefronts converge to a focal spot. **b-c,** Local phase (in degrees) of each element as a function of the element parameter and the frequency. **b** refers to unit cells with period 1.8 µm, 1 µm width and variable lengths. **c** refers to unit cells with period 1.2 µm, 9 µm length and variable widths. **d,** Designed phase profile of the two metalenses (see Supplementary Information for additional information). The design frequency was 1452 cm$^{-1}$ and 1448 cm$^{-1}$ for Metalens 1 and Metalens 2 respectively. **e,** Optical images of the written metalenses. Metalens 1 was written, characterised, erased and subsequently the same area was reconfigured into Metalens 2. **f,** s-SNOM image of Metalens 1 showing focusing of polaritons at 1452 cm$^{-1}$. **g,** s-SNOM image of Metalens 2 at 1445 cm$^{-1}$. **h,** Intensity profile (square of s-SNOM signal) at the cross-section of the diffraction-limited focused polariton beam for both lenses (dashed vertical lines in panels **f** and **g**). Scale bars are 5 µm.

s-SNOM characterisation was carried out after each step and clearly reveals the focusing effect of the two designed lenses (Figures 4f and 4g). Figure 4h shows the confinement of polaritons at the focal spot. Quantitative analysis (see Supplementary Information) of both metalenses confirms that we were able to achieve spherical aberration free, diffraction-limited focusing using both approaches: the lateral size of the focal spots are 1.6 $\mu$m and 2 $\mu$m respectively, which is in good agreement with the respective Abbe limits (1.66 $\mu$m and 1.90 $\mu$m respectively). The focal spots are 23 % and 29 % of the free space wavelength respectively.

In summary, our results clearly establish that the hBN-GST heterostructure used in this work can serve as a versatile platform to arbitrarily control polaritons at the nanoscale to achieve freeform, transformation, and meta optics[18]. While we chose to integrate hBN with GST, it can be readily combined with other polaritonic vdW materials[3] and other phase change materials[31], thereby enabling a whole range of deeply subwavelength polaritonic devices, from visible to infrared spectral regimes. Fully-fledged polaritonic circuits can be cheaply fabricated without the need for traditional photolithography, allowing the low-cost realisation of biosensors[15], and high-density optical storage, which benefit from the extreme volume confinement that can be achieved with polaritons. Additionally, the reconfigurability offered by GST-vdW heterostructures and the possibility of electrically switching GST – vdW materials heterostructures will pave the way to applications such as modulators, photo-detectors and, more generally, programmable optical devices as optical counterparts to field programmable gate arrays.

## Methods

**Sample Fabrication**: A 55 nm GST-326 film (with 15 nm of ZnS:SiO$_2$ protection layer) was sputter-coated onto a 1 mm thick CaF$_2$ substrate. Lithography was performed to define alignment markers (positive tone photoresist S1813 spin coated at 3000 RPM and baked at 115 °C for 90 seconds) followed by Pt sputtering (30 nm) and lift-off at room temperature. Isotopically pure h$^{11}$BN flakes were mechanically exfoliated onto the substrate after plasma activation (5 minutes of $O_2$ plasma at 100 W) using a standard Scotch tape process. We removed traces of the glue from the Scotch tape by placing the sample in acetone for 10 minutes followed by an isopropyl alcohol rinse for 5 minutes and drying with nitrogen. Afterwards, samples were further cleaned with $O_2$ plasma (10 minutes of $O_2$ plasma at 100 W). The thicknesses of the flakes and of GST were confirmed through AFM measures (Cypher AFM from Asylum Research).

**Reconfigurable Pattern Writing**: see Supplementary Method 3 for details on the technique used to write and erase patterns in GST.

**Lens and metalens parameters**: see Supplementary Method 2 for a summary of the fabricated lens parameters.

**Numerical Simulations**: Numerical simulations were performed using Lumerical Mode solutions and Lumerical FDTD. 1D simulations in Mode Solutions were used to calculate the effective indices $n_{\text{eff,a}}$, $n_{\text{eff,c}}$. A mesh size of 1 nm was used to compute the fundamental mode profile and effective indices of the hBN-GST-326 heterostructures in the RS2 band of h[11]BN. Here, hBN was modelled as an anisotropic dielectric with its permittivity values obtained from the Lorentz model (Supplementary Method 1). The effective indices of the waveguides were calculated via 2D simulations of their cross-sections, also done in Modes Solutions. Metalenses were designed by first simulating numerically each element in a periodic environment with full wave 3D simulations performed in FDTD. The result was a library of elements, used to implement the required phase profile. The final metalens was also completely simulated to verify the focusing behaviour. For 3D simulations we used a mesh size smaller than 50 nm along *x*- and *y*-axes and 5 nm along the *z*-axis. A series of dipole sources were used to excite the polariton modes in hBN in the RS2 band.

**s-SNOM Measurements**: The near field scans were obtained using a commercial system from NeaSpec. Tapping-mode AFM is used (tapping amplitude of 130 nm, Pt-Ir coated tips with resonant frequency of ~300 kHz, tip diameter of ~20 nm). A quantum cascade laser (QCL) array from Daylight Solutions was used as a tuneable mid-IR source for imaging. The phase-amplitude images are obtained using a pseudo-heterodyne demodulation. The images of lenses and metalenses were processed using the technique shown in the Supplementary Method 4 to isolate the polaritons focused by the structure. In all cases polaritons are self-launched by the edges[32]. See Supplementary method 5 for additional measurements.

## **Acknowledgements**

Funding: This work was supported by the NSF EFRI, award no. 1542807. This work was performed in part at the Center for Nanoscale Systems (CNS), a member of the National Nanotechnology Coordinated Infrastructure Network (NNCI), which is supported by the National Science Foundation under NSF award no. 1541959. M.T. acknowledges the support of the Swiss National Science Foundation (SNSF) grant no. 168545 and 177836. S.L.O. acknowledges "Fondazione Angelo Della Riccia", and the program for "International Mobility of Researchers" of the University of Naples "Federico II" (Italy), for financial support. The h$^{11}$BN crystal growth was supported by the National Science Foundation, award number 1538127. K.W. and T.T. acknowledge support from the Elemental Strategy Initiative conducted by the MEXT, Japan and and the CREST (JPMJCR15F3), JST.

## **Author Contributions**

K.C., M.T. and X.Y. conceived the project. M.T., K.C., X.Y., C.S., A.A. and F.C. devised experiments. X.Y., C.S. designed and implemented the setup for writing patterns on GST-326 with help from R.L. and N.A.R.. J.L. and J.E. provided isotopically pure h$^{11}$BN for final devices. C.P. and M.W. optimized and deposited the GST layer on the substrate. K.W., T.T., L.A.J. and P.K. provided natural abundance hBN for initial tests. X.Y. and C.S. wrote and erased patterns on h$^{11}$BN/GST-326 heterostructures. K.C., M.T., S.L.O., C.S. and A.A., performed s-SNOM and AFM scans. K.C., M.T., X.Y. and C.S. performed FDTD simulations. M.T. and K.C. analysed the experimental data. F.C. led the project. All authors contributed to discussions and manuscript preparation.

## **Author Information**

The Authors declare no competing financial interests. Please contact M.T. for any request of materials and additional data.